\documentclass[manuscript]{acmart}

\usepackage{graphicx}
\usepackage{subcaption}
\usepackage{cleveref}
\everypar\expandafter{\the\everypar\loosness=-1}
\AtBeginDocument{%
  }

\setcopyright{acmlicensed}
\copyrightyear{2024}
\acmYear{2024}
\acmDOI{XXXXXXX.XXXXXXX}

\begin{document}

\title{The Role of Fake Users in Sequential Recommender Systems}

\titlenote{This work was partially supported by projects FAIR (PE0000013) and SERICS (PE00000014) under the MUR National Recovery and Resilience Plan funded by the European Union - NextGenerationEU. Supported also by the ERC Advanced Grant 788893 AMDROMA,  EC H2020RIA project “SoBigData++” (871042), PNRR MUR project  IR0000013-SoBigData.it. This work has been supported by the project NEREO (Neural Reasoning over Open Data) project funded by the Italian Ministry of Education and Research (PRIN) Grant no. 2022AEFHAZ.}
\author{Filippo Betello}
    \orcid{0009-0006-0945-9688}
    \affiliation{
      \institution{Sapienza University of Rome}
      \city{Rome}
      \country{Italy}
      }
    \email{betello@diag.uniroma1.it}
\acmArticleType{Review}


\begin{abstract}
Sequential Recommender Systems (SRSs) are widely used to model user behavior over time, yet their robustness remains an under-explored area of research. In this paper, we conduct an empirical study to assess how the presence of fake users —who engage in random interactions, follow popular or unpopular items, or focus on a single genre —impacts the performance of SRSs in real-world scenarios. We evaluate two SRS models across multiple datasets, using established metrics such as Normalized Discounted Cumulative Gain (NDCG) and Rank Sensitivity List (RLS) to measure performance. While traditional metrics like NDCG remain relatively stable, our findings reveal that the presence of fake users severely degrades RLS metrics, often reducing them to near-zero values. These results highlight the need for further investigation into the effects of fake users on training data and emphasize the importance of developing more resilient SRSs that can withstand different types of adversarial attacks.

\end{abstract}



\maketitle

\section{Introduction}
Recommender Systems (RSs) have become an essential part of our daily lives, helping users navigate the vast online information landscape \cite{adomavicius2005toward}. With the global expansion of e-commerce services, social media platforms and streaming services, these systems have become essential for personalising content delivery and increasing user engagement \cite{zhang2019deep}. 

Over the last several years, Sequential Recommender Systems (SRSs) have gained significant popularity as an effective method for modeling user behavior over time \cite{quadrana2018sequence}. By capitalizing on the temporal dependencies within users’ interaction sequences, these systems can make more precise predictions about user preferences \cite{wang2019sequential}. This approach allows for a more nuanced understanding of user behavior, leading to recommendations that are better tailored to individual needs and preferences. As a result, SRSs have become a critical component in various applications, ranging from e-commerce \cite{hwangbo2018recommendation} to music recommendation \cite{afchar2022explainability}, where understanding and anticipating user preferences is key to enhancing user experience and engagement.

In recent years, the prevalence of bots (fake users) on social media platforms has increased dramatically \cite{ferrara2016rise}. It is estimated that Amazon, for example, spends 2\% of its net revenue each year fighting counterfeiting \cite{Daniels_2024}. While several techniques have been identified to counteract this growing problem \cite{mendoza2020bots, mazza2019rtbust}, a detailed investigation in the area of sequential recommendation systems is still lacking. \citet{li2022revisiting} aims to fill this gap by investigating the impact of bot-generated data on sequential recommendation models. Specifically, it seeks to determine an optimal bot-generation budget and analyze its impact on popular matrix factorization models. Indeed, controlling and maintaining a large number of bots is costly. 

Therefore, it is possible to create a limited number of bots that can significantly influence the prominence of a particular item or category. By strategically deploying these bots, the visibility and perceived importance of the targeted item or category can be enhanced, making it stand out more compared to others. Imagine if, by using fake users, it were possible to raise the profile of a certain category or product or, conversely, to lower the profile of another. This scenario represents a form of unfair competition and is therefore crucial to study. Understanding how fake users behave in controlled environments allows us to assess their impact on real users. It is also important to investigate whether partially coordinated fake users can actively improve the performance or predictions of a particular category or item. 

In this paper, we investigate the impact of fake users on sequential recommendation systems. Specifically, we investigate how the inclusion of a certain percentage of bots affects the performance of real users. These bots are programmed to deal with random items, popular items, unpopular items and items within the same category. 

Our experiments focus on the following research questions:
\begin{itemize}
    \item \textbf{RQ1}: How does the value of standard metrics such as NDCG change for real users depending on the type and increasing number of fake users?
    \item \textbf{RQ2}:How do recommendation lists for real users differ from those generated without fake users?
    \item \textbf{RQ3}: Are more or less popular items favoured by the presence of fake users with certain types of interactions?
\end{itemize}

We evaluate our hypothesis using two different models, SASRec \cite{kang2018self} and GRU4Rec \cite{hidasi2016sessionbased}, and by employing four different datasets, namely MovieLens 1M, MovieLens 100k \cite{harpermovielens}, Foursquare New York City and Foursquare Tokyo \cite{yang2014modeling}.

\section{Related Work}

\subsection{Sequential Recommender Systems}
Sequential recommendation systems (SRSs) use algorithms that analyze a user's past interactions with items to provide personalized recommendations over time. These systems have found widespread application in areas such as e-commerce \cite{schafer2001commerce,hwangbo2018recommendation}, social media \cite{guy2010social,amato2017recommendation}, and music streaming services \cite{schedl2015music,schedl2018current,afchar2022explainability}. Unlike traditional recommender systems, SRSs take into account the sequence and timing of user interactions, resulting in more precise predictions of user preferences and behaviors \cite{wang2019sequential}.

Various methods have been developed to implement SRSs. Early approaches used Markov Chain models \cite{fouss2005novel,fouss2005web}, which, despite their simplicity, struggled with capturing complex dependencies in long-term sequences. More recently, Recurrent Neural Networks (RNNs) have become prominent in this domain \cite{donkers2017sequential,hidasi2016sessionbased,quadrana2017personalizing}. RNNs encode a user's historical preferences into a vector that is updated at each time step to predict the next item in the sequence. However, RNNs can encounter difficulties with long-term dependencies and generating diverse recommendations.

The attention mechanism \cite{vaswani2017attention} has introduced another promising approach. Models like SASRec \cite{kang2018self} and BERT4Rec \cite{sun2019bert4rec} leverage this mechanism to dynamically weight different parts of the sequence, capturing key features to enhance prediction accuracy. 

Additionally, Graph Neural Networks have recently gained traction in the recommendation system field, particularly within the sequential domain \cite{chang2021sequential,fan2021continuous}. These networks excel at modeling complex relationships and dependencies, further advancing the capabilities of SRSs \cite{wu2022graph,purificato2023sheaf}.

\subsection{Training Perturbations}
Robustness is an important aspect of SRSs as they are vulnerable to noisy and incomplete data. 
\cite{betello2024investigating, oh2022rank} investigated the effects of removing items at the beginning, middle and end of a sequence of temporally ordered items and found that removing items at the end of the sequence significantly affected all performances.

\citet{yin2024poisoning} design an attacker-chosen targeted item in federated recommender systems without requiring knowledge about user-item rating data, user attributes, or the aggregation rule used by the server. 
While studies are being conducted in other areas of recommendation \cite{10264111, 10.1145/3539618.3592070} and several techniques have been identified to counteract this growing problem \cite{mendoza2020bots,mazza2019rtbust}, a detailed investigation in the area of sequential recommendation systems is still lacking.

\citet{li2022revisiting} aim to address this issue by examining how bot-generated data affects sequential recommendation models. Their research focuses on finding the optimal budget for bot generation and assessing its influence on widely used matrix factorization models. Indeed, controlling and maintaining a large number of bots is costly. Previous research has proposed attacks using a limited number of users and clustering models \cite{wang2024clusterpoison}, but these have not been extensively studied in the context of sequential recommendations.

To the best of our knowledge, our research is completely novel and breaks new ground. It explores the role that fake users might play in influencing real users. This study aims to shed light on the potential impact that fake users could have on the behaviour, opinions and interactions of real users within sequential recommendation systems. 

\section{Methodology}
\subsection{Background}
The main objective of sequential recommendation systems is to predict the user's next interaction in a given sequence. Suppose we have a set of $n$ users, represented as $\mathcal{U} \subset \mathbb{N^+}$, and a corresponding set of $n$ items, represented as $\mathcal{I} \subset \mathbb{N^+}$. Each user $u \in \mathcal{U}$ is associated with a time-ordered sequence of interactions $S_u = [s_1, \dots, s_{L_u}]$, where each $s_i \in \mathcal{I}$ denotes the $i$-th item with which the user has interacted. The length of this sequence, $L_u$, is greater than 1 and varies from user to user.

A sequential recommendation system (SRS), denoted $\mathcal{M}$, processes the sequence up to the $L$-th item, denoted $S_u^L = [s_1, \dots, s_L]$, to suggest the next item, $s_{L+1}$. The recommendation output, $r^{L+1} = \mathcal{M}(S_u^L) \in \mathbb{R}^m$, is a score distribution over all possible items. This distribution is used to create a ranked list of items, predicting the most likely interactions for user $u$ in the next step, $L+1$.

\subsection{Fake user design}
Given that each item in the set $\mathcal{I}$ has a popularity value determined by user interactions, we designed four types of fake user scenarios:
\begin{itemize}
    \item \textbf{Random}: Items are randomly sampled from the entire set $\mathcal{I}$. Formally, each item $s_i$ in the sequence $S_u$ is selected with probability $\frac{1}{|\mathcal{I}|}$. 
    \item \textbf{Popularity}: Items are sampled according to a popularity-based probability distribution $P_{\text{pop}}$, where the probability of selecting item $s_i$ is proportional to its popularity $p_i$.
    \item \textbf{Unpopularity}: Similar to the popularity-based scenario, but with a distribution $P_{\text{unpop}}$ that inversely favors popular items. Here, the probability of selecting item $s_i$ is inversely proportional to its popularity, $\Pr(s_i) \propto \frac{1}{p_i}$, favoring less popular items.
    \item \textbf{Genre}: In this scenario, items are sampled exclusively from a specific genre. It is only applied to the ML datasets.

\end{itemize}

These fake users sequences will contain unique items to ensure there are no repetitions. While the first scenario involves users acting independently without any sense of cooperation, the middle two scenarios introduce a level of implicit cooperation. Specifically, users in these scenarios tend to converge on viewing either highly popular or highly unpopular items, reflecting a collective behavior. The average length of the sequences will be the same as that of real users. The proportion of synthetic users will vary, comprising 1\%, 5\%, 10\%, 15\% and 20\% of the original dataset. The fake users are only used in the training data, leaving the test data unaffected.

\subsection{Models}
In our study, we use two different architectures to validate our results:

\begin{itemize}
\item \textbf{SASRec} \cite{kang2018self}, which uses self-attention mechanisms to evaluate the importance of each interaction between the user and the item.

\item \textbf{GRU4Rec} \cite{hidasi2016sessionbased}, a RNN model that uses gated recurrent units (GRUs) \cite{cho2014learning} to improve prediction accuracy.
\end{itemize}

We chose these two models because they have demonstrated exceptional performance in numerous benchmarks and are widely cited in the academic literature.
Moreover, since one model is based on attention mechanisms and the other on RNNs, their different network operations make it particularly interesting to evaluate their behaviour.

\subsection{Datasets}

\begin{table}[t!]
    \caption{Dataset statistics after pre-processing; users and items not having at least 5 interactions are removed. Avg. and Med. refer to the Average and Median of $\frac{\mathrm{Actions}}{\mathrm{User}}$, respectively.}
      \begin{tabular}{l||ccc|ccc}
        \toprule
        Name & Users & Items & Interactions &  Density & Avg. & Med. \\ 
        \midrule
        FS-NYC & 1,083 & 9,989 & 179,468 & 1.659 & 165 & 116\\
        FS-TKY & 2,293 & 15,177 & 494,807 & 1.421 & 215 & 146\\
        ML-100k & 943 & 1,349 & 99,287 & 7.805 & 105 & 64 \\
        ML-1M &  6,040 & 3,416 & 999,611 & 4.845 & 165 & 96\\
        \bottomrule
      \end{tabular}
\label{tab:dataset_info}
\end{table}

We use four different datasets:

\textbf{MovieLens} \cite{harpermovielens}: Frequently utilized to evaluate recommender systems, this benchmark dataset is employed in our study using both the 100K and 1M versions.

\textbf{Foursquare} \cite{yang2014modeling}: This dataset includes check-in data from New York City and Tokyo, collected over a span of roughly ten months.

The statistics for all the datasets are shown in \Cref{tab:dataset_info}. Our pre-processing technique adheres to recognised principles, such as treating ratings as implicit, using all interactions without regard to the rating value, and deleting users and things with fewer than 5 interactions \cite{kang2018self, sun2019bert4rec}.
For testing, as in \cite{sun2019bert4rec, kang2018self}, we keep the most recent interaction for each user, while for validation, we keep the second to last action. The remaining interactions are added to the training set, which is the only one affected by the fake users perturbation.

We focus exclusively on genres in the ML dataset, as it is the only dataset that contains category information. Specifically, we select only those categories that represent more than 5\% of the total items in the dataset.

\subsection{Evaluation}
\looseness -1  We only carry out the evaluation on the part of the real users. To evaluate the performance of the models, we employ traditional evaluation metrics used for Sequential Recommendation: Precision, Recall, MAP and NDCG. Moreover, to investigate the stability of the recommendation models, we employ the Rank List Sensitivity (RLS) \cite{oh2022rank}: it compares two lists of rankings $\mathcal{X}$ and $\mathcal{Y}$, one derived from the model trained under standard conditions  and the other derived from the model trained with perturbed data.

Given these two rankings, and a similarity function $sim$ between them, we can formalise the RLS measure as

\begin{equation}
    \boldsymbol{\mathrm{RLS}} = \frac{1}{|\mathcal{X}|} \sum\limits_{k=1}^{|\mathcal{X}|} \text{sim}(R^{X_k}, R^{Y_k})
\end{equation}

where $X_k$ and $Y_k$ represent the $k$-th ranking inside $\mathcal{X}$ and $\mathcal{Y}$ respectively.

RLS's similarity measure can be chosen from two possible options:
\begin{itemize}
    \item \looseness -1\textbf{Jaccard Similarity (JAC)} \cite{jaccard1912distribution} is a normalized measure of the similarity of the contents of two sets. A model is stable if its Jaccard score is close to 1.
    \begin{equation}
       \boldsymbol{\mathrm{JAC(X,Y)}} = \! \frac{|X \cap Y|}{|X \cup Y|}
    \end{equation}
    \item \textbf{Finite-Rank-Biased Overlap (FRBO)} \cite{betello2024investigating} measures the similarity of orderings between two rank lists. Higher values indicate that the items in the two lists are arranged similarly:
    \begin{equation*}\label{eq:FRBOsimple}
    \boldsymbol{\mathrm{FRBO(X,Y)@k}} = \frac{1-p}{1-p^k}\sum_{d=1}^{k} p^{d-1} \frac{|X[1:d] \cap Y[1:d]|}{d}
\end{equation*}
\end{itemize}
All metrics are computed ``@$k$'', meaning that we use just the first $k$ recommended items in the output ranking, with $k \in \{10,20\}$.

\subsection{Experimental Setup}
All experiments were performed on a single NVIDIA RTX A6000 with 10752 CUDA cores and 48 GB of RAM. We train the models for 500 epochs, fixing the batch size to 128 and by using the Adam optimizer \cite{kingma2017adam} with a lr of $10^{-3}$. To run our experiments, we use the EasyRec library \cite{betello2024reproducible}.

\begin{figure}[t]
    \centering
    
    \begin{subfigure}[b]{0.45\textwidth}
        \centering
        \includegraphics[width=\textwidth]{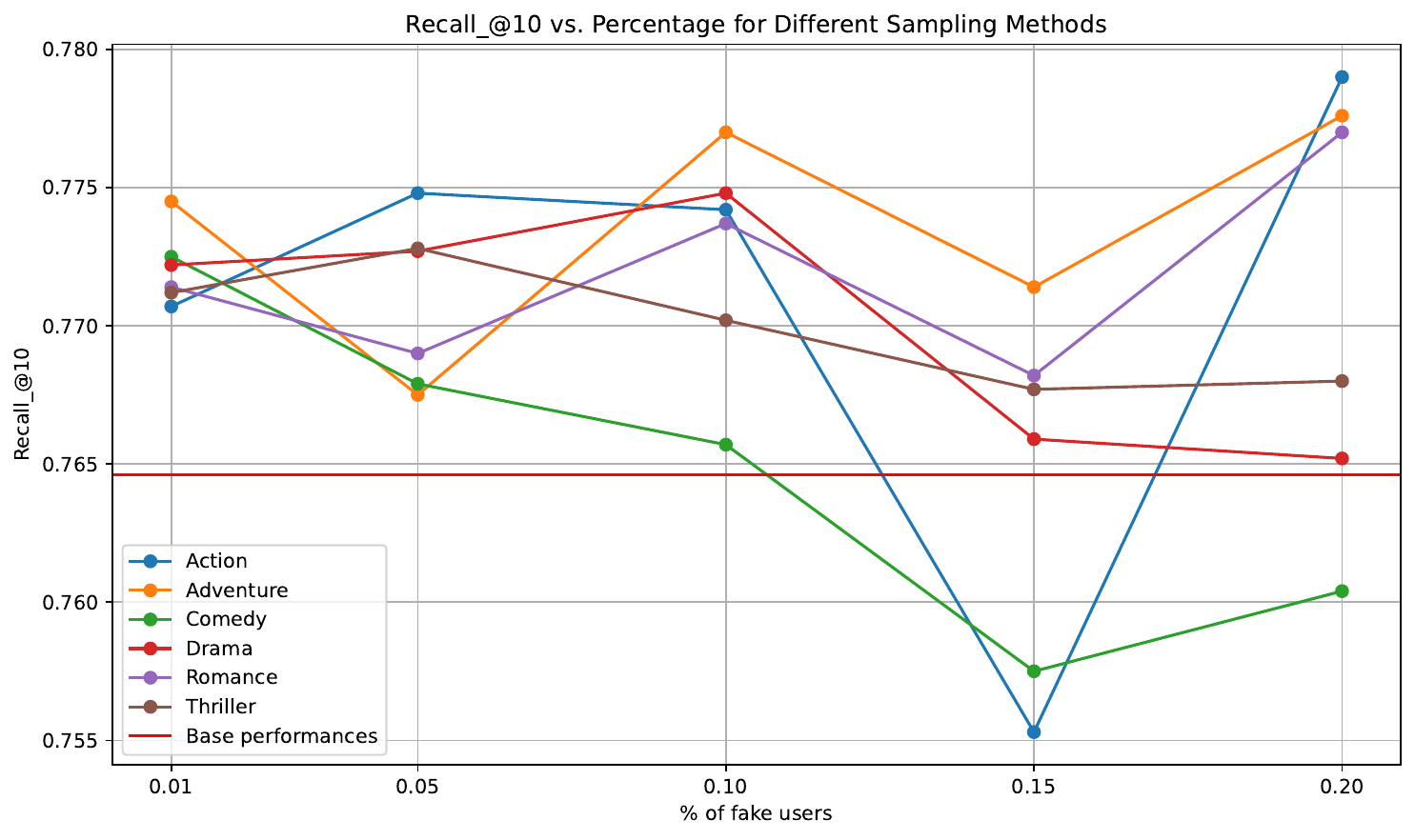} 
        \caption{NDCG@20 ML-1M SASRec}
        \label{fig:plot1}
    \end{subfigure}
    \hfill
    \begin{subfigure}[b]{0.45\textwidth}
        \centering
        \includegraphics[width=\textwidth]{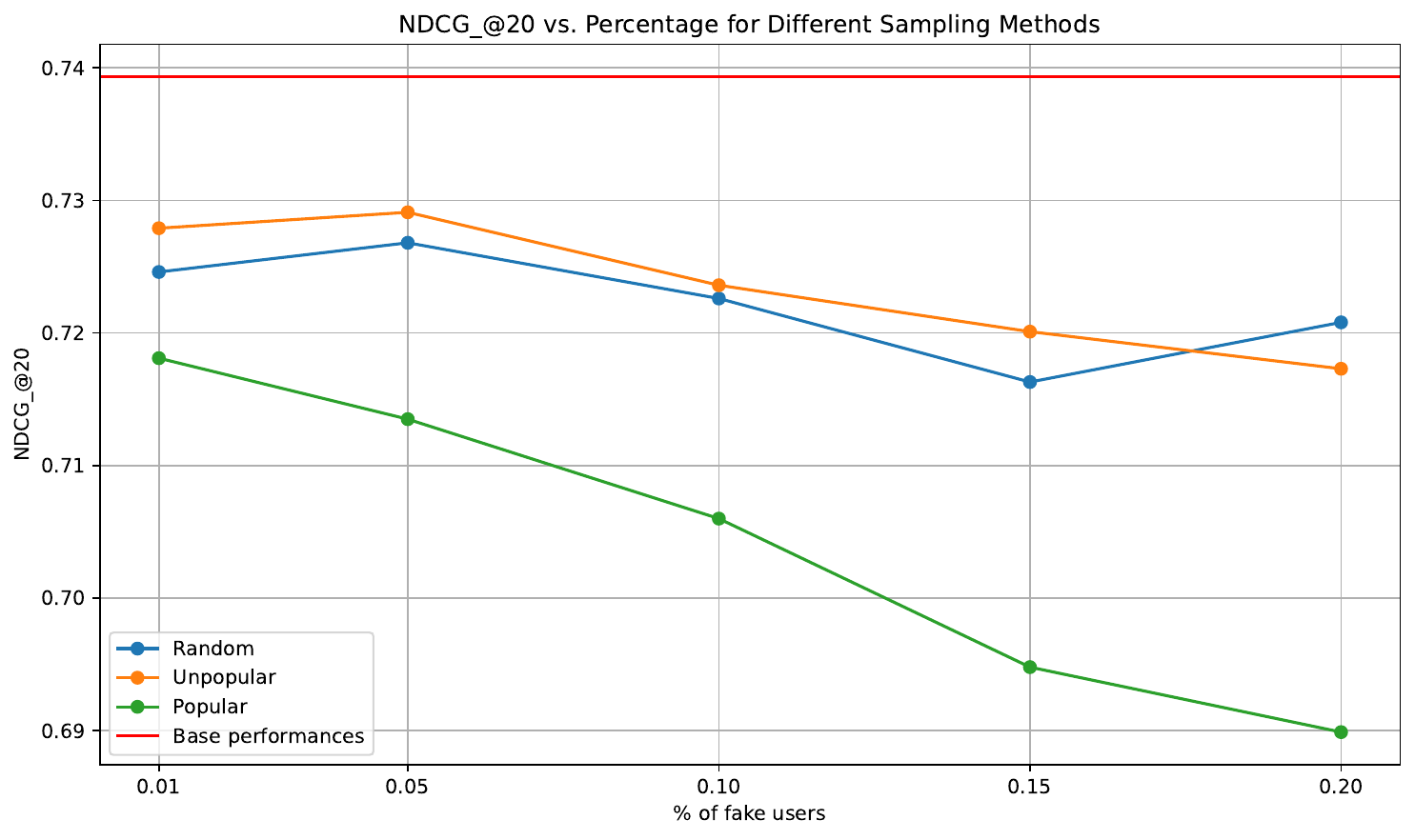} 
        \caption{MAP@10 FS-TKY GRU4Rec}
        \label{fig:plot2}
    \end{subfigure}

    \begin{subfigure}[b]{0.45\textwidth}
        \centering
        \includegraphics[width=\textwidth]{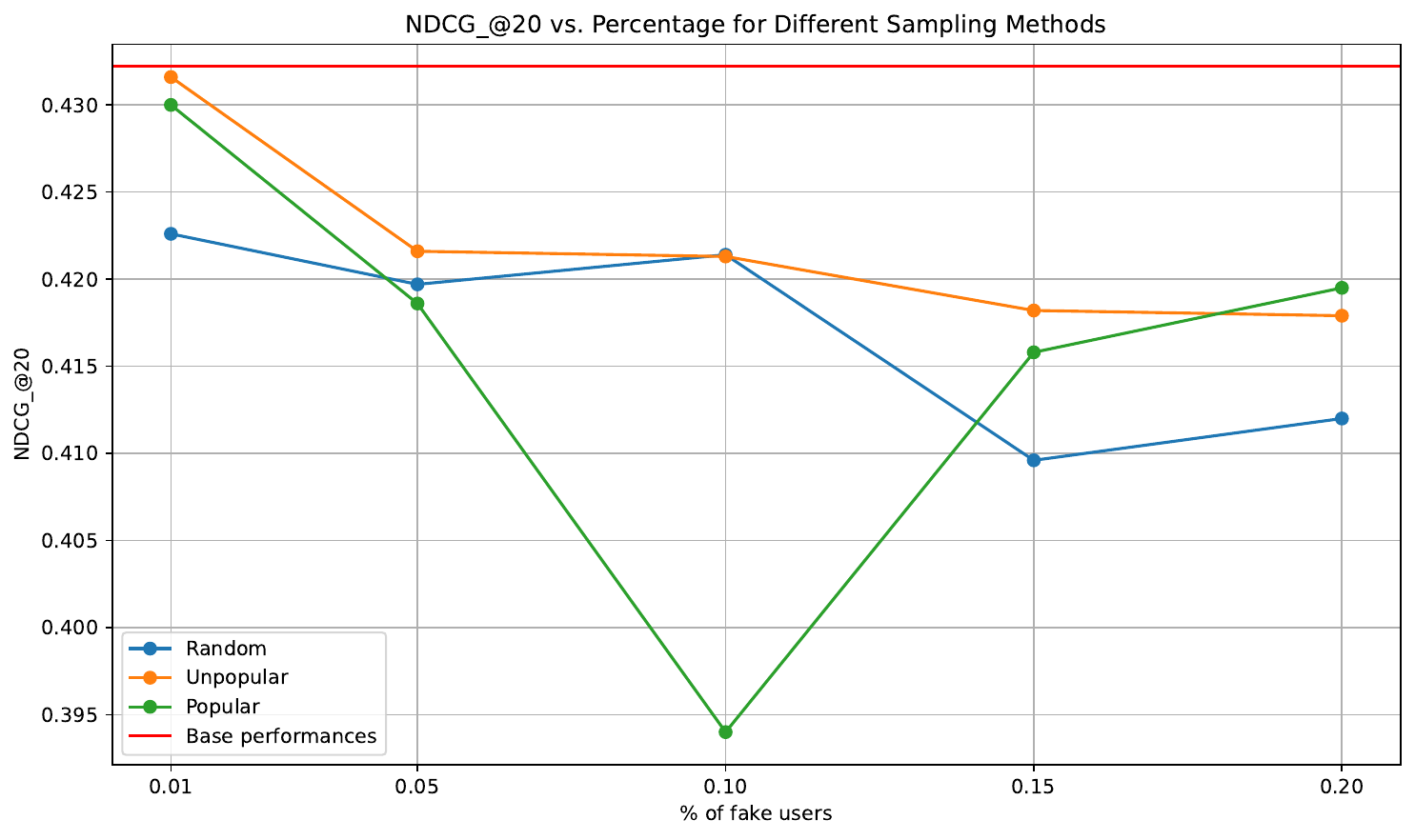} 
        \caption{NDCG@20 ML-100k GRU4Rec}
        \label{fig:plot3}
    \end{subfigure}
    \hfill
    \begin{subfigure}[b]{0.45\textwidth}
        \centering
        \includegraphics[width=\textwidth]{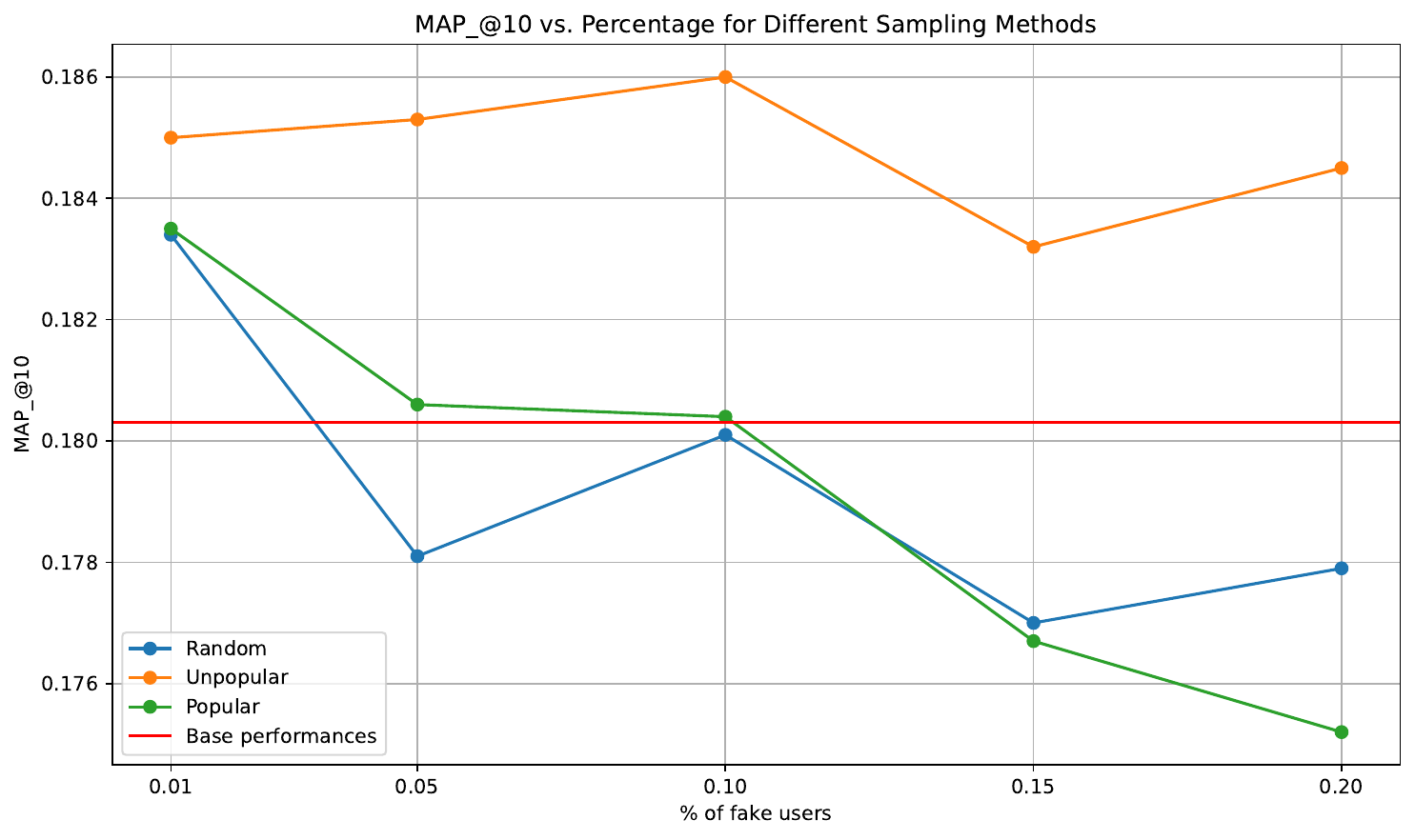} 
        \caption{NDCG@20 FS-NYC SASRec}
        \label{fig:plot4}
    \end{subfigure}
    
    \caption{Plots of various metrics for all the datasets considered as the percentage of fake users increases. The baseline is shown as a horizontal solid line, while other lines show the metrics as the percentage of fake users changes for the three scenarios considered.}
    \label{fig:four_plots}
\end{figure}

\section{Results}
Our experiments aim to address the following research questions:

\begin{itemize}
    \item \textbf{RQ1}: How does the value of standard metrics such as NDCG change for real users depending on the type and increasing number of fake users?
    \item \textbf{RQ2}:How do recommendation lists for real users differ from those generated without fake users?
    \item \textbf{RQ3}: Are more or less popular items favoured by the presence of fake users with certain types of interactions?
\end{itemize}
\subsection{RQ1: Impact of Fake Users on Standard Metrics for Real Users}

In \Cref{fig:four_plots} the results for all datasets considered are shown for both models using the standard metrics.

Regarding the SASRec shown in \Cref{fig:plot4} for the FS-NYC dataset, we observe that the performance tends to improve slightly for the unpopular scenario for the NDCG@20 metric, while for the popular and random interaction there is a gradual but consistent decline in performance. Regarding genre interactions in the ML-1M dataset, shown in \Cref{fig:plot1}, all genres appear to positively impact the NDCG metric. A more detailed analysis using RLS metrics is presented in \Cref{sec:RQ2}.

In the case of GRU4Rec \cref{fig:plot2,fig:plot3}, there is a slow but steady decline in performance for the ML-100k and FS-TKY datasets, with the decline occurring in a predictable manner for both metrics considered, as the percentage of fake users increase.

\subsection{RQ2: Analysis of Recommendation Lists Generated for Real Users} \label{sec:RQ2}
\begin{figure}[h]
    \centering
    
    \begin{subfigure}[b]{0.45\textwidth}
        \centering
        \includegraphics[width=\textwidth]{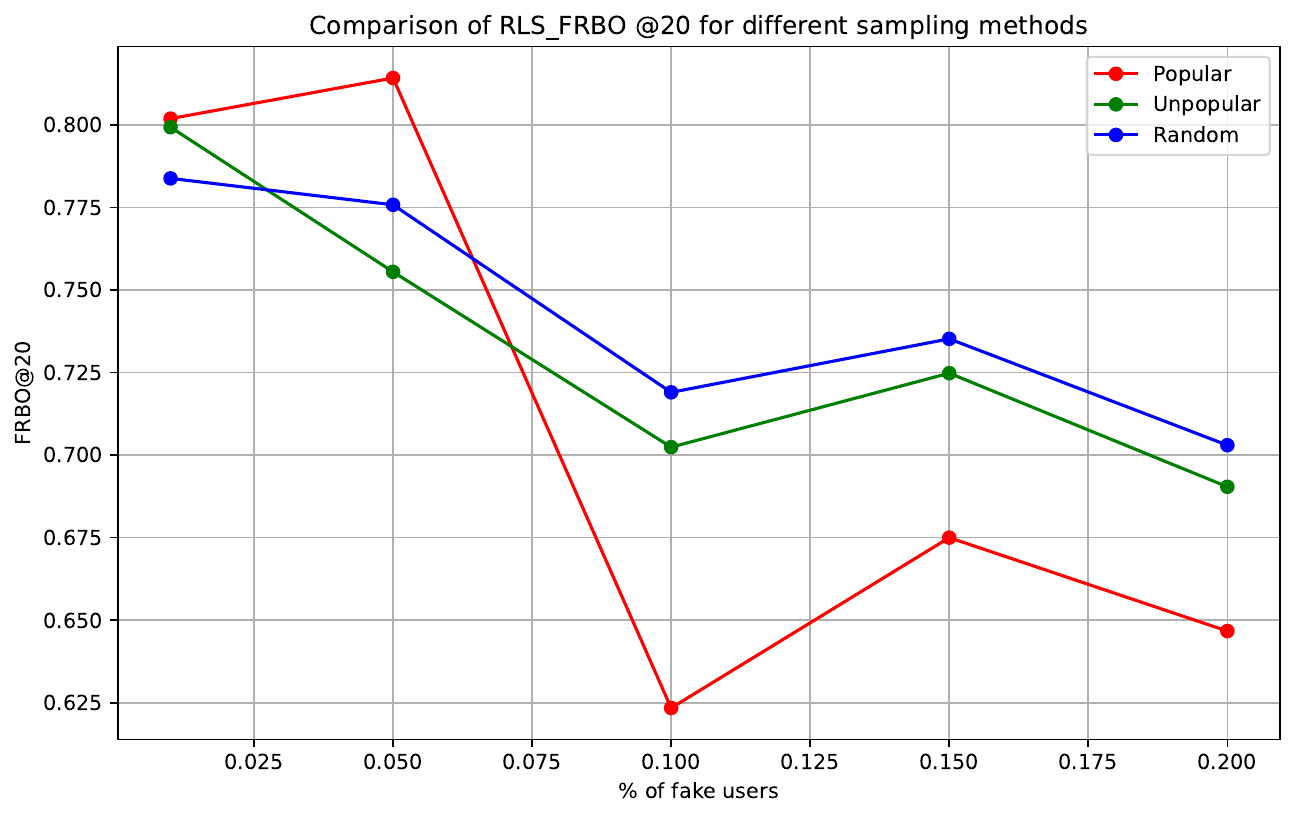} 
        \caption{RLS-FRBO ML100k SASRec}
        \label{fig:plot1_rls}
    \end{subfigure}
    \hfill
    \begin{subfigure}[b]{0.45\textwidth}
        \centering
        \includegraphics[width=\textwidth]{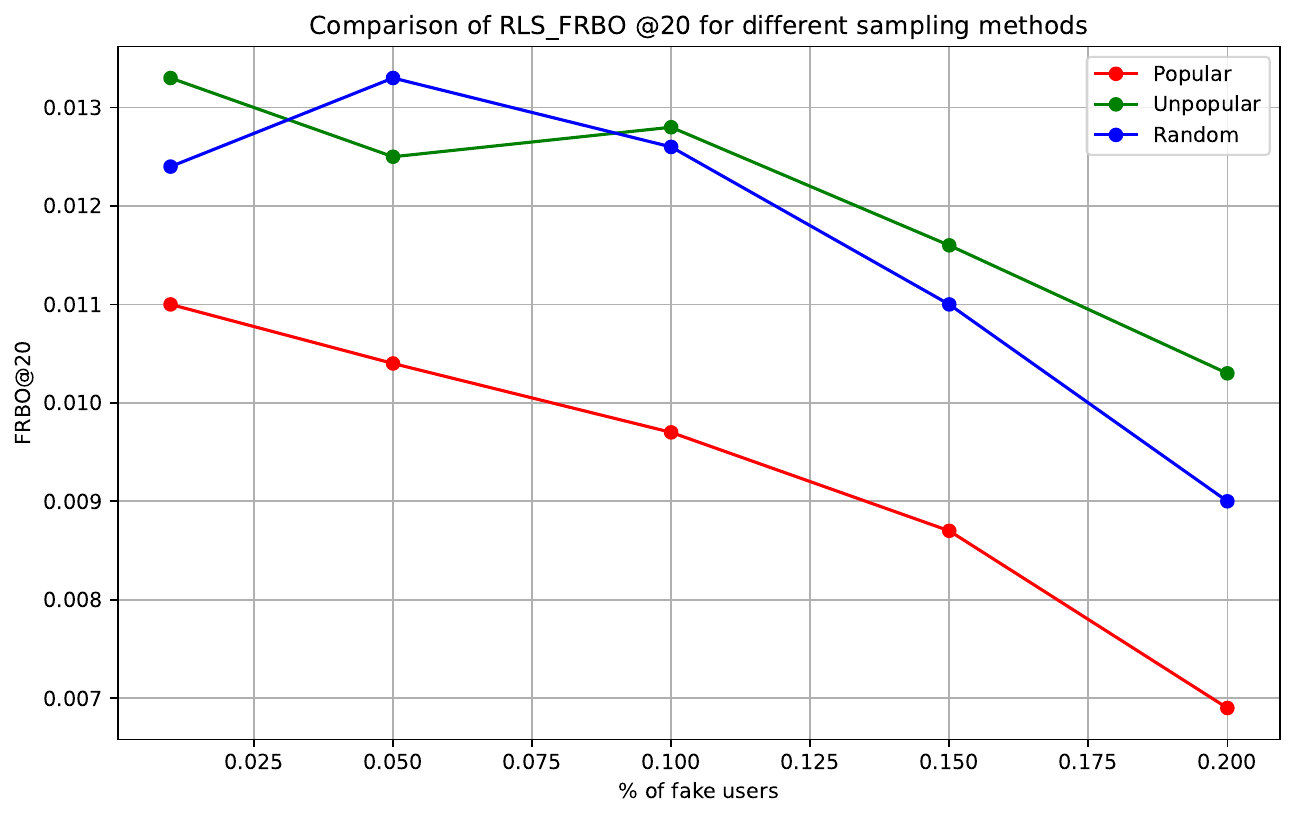} 
        \caption{RLS-FRBO FS-TKY SASRec}
        \label{fig:plot2_rls}
    \end{subfigure}

    \begin{subfigure}[b]{0.45\textwidth}
        \centering
        \includegraphics[width=\textwidth]{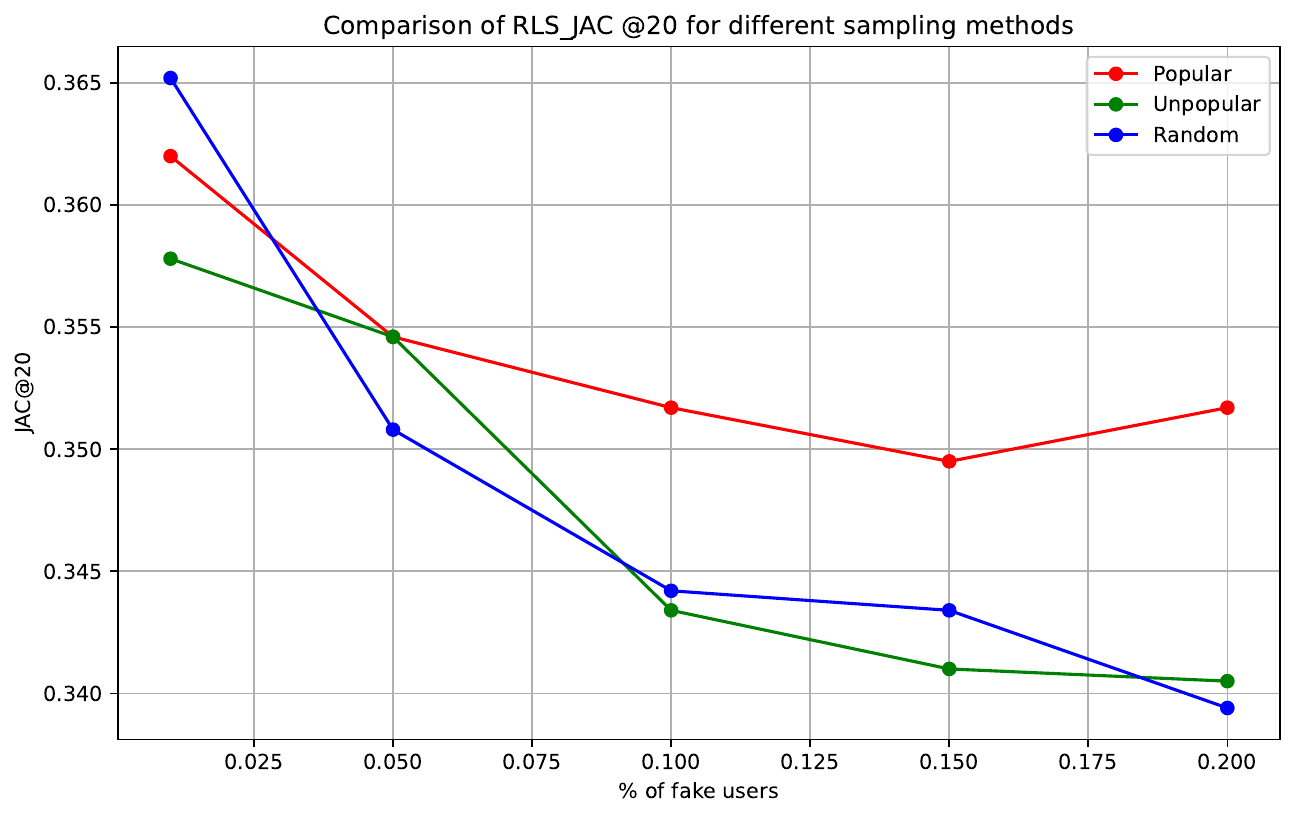} 
        \caption{RLS-JAC ML-1M GRU4Rec}
        \label{fig:plot3_rls}
    \end{subfigure}
    \hfill
    \begin{subfigure}[b]{0.45\textwidth}
        \centering
        \includegraphics[width=\textwidth]{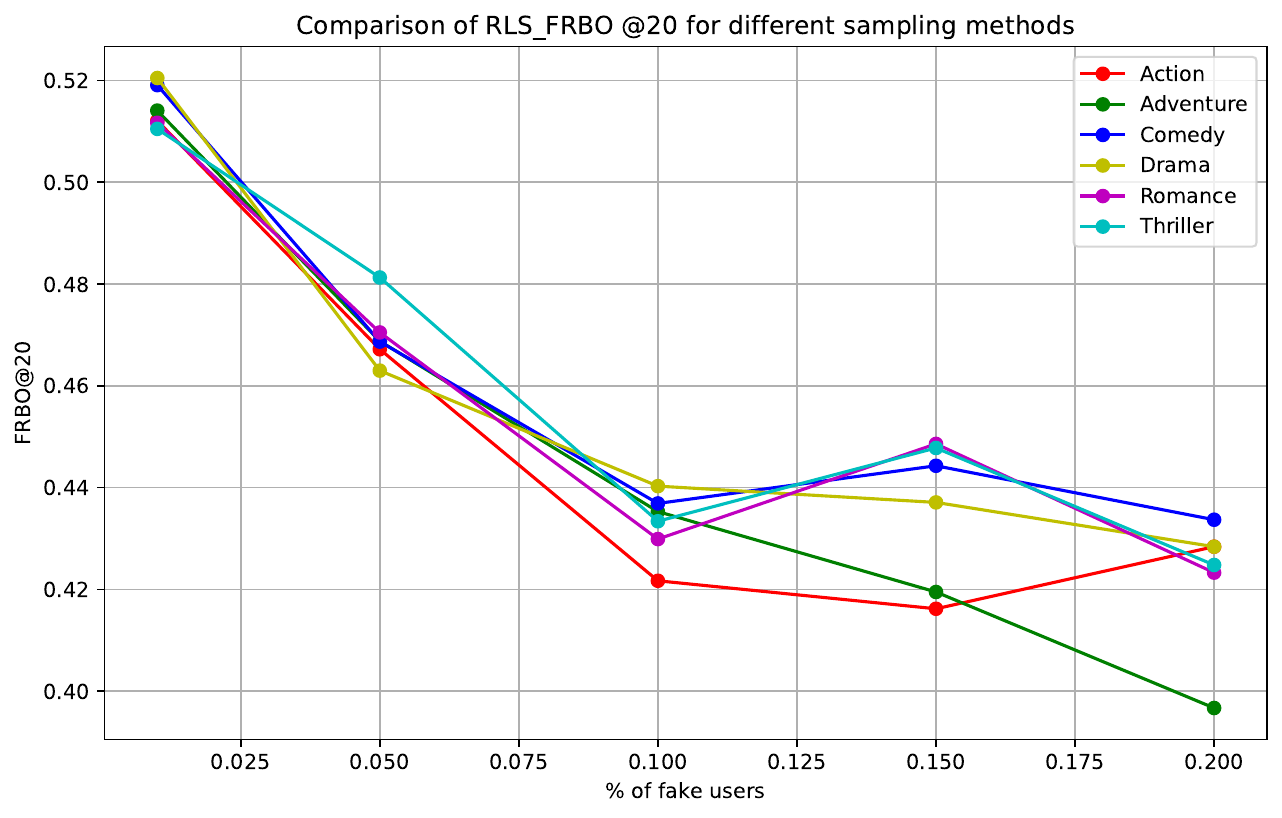} 
        \caption{RLS-FRBO ML-100k GRU4Rec}
        \label{fig:plot4_rls}
    \end{subfigure}
    
    \caption{Plots of RLS metrics for all the datasets considered as the percentage of fake users increases. The metrics are shown as the percentage of fake users changes for the three scenarios considered.}
    \label{fig:four_plots_rls}
\end{figure}

In \Cref{fig:four_plots_rls} we present the RLS metrics for all datasets considered, comparing the performance of the two models. These metrics are derived from predictions made by the standard model - without fake users - and predictions made after training with fake users.

When analysing the SASRec model on the ML-100k dataset (\cref{fig:plot1_rls}), SASRec shows minimal performance degradation. Conversely, the FS-TKY dataset gives less favourable results, with significantly worse performance and a Jaccard index close to 0, indicating that the generated lists have almost no overlap with the original lists (\cref{fig:plot2_rls}).

Figures \Cref{fig:plot3_rls,fig:plot4_rls} show the performance on the ML-100k dataset for genre sampling and the ML-1M dataset for the other sampling methods. On the ML-1M dataset, the performance is relatively good, although the Jaccard index remains low at around 0.35 (\cref{fig:plot3_rls}). For ML-100k and genre interactions, the degradation in performance is consistent across all genres, with the degradation worsening as the number of fake users increases.

The evaluation metrics for Foursquare show a significant drop in performance compared to other datasets, highlighting the limitations of the dataset \cite{klenitskiy2024does}.

An additional observation is that as the number of fake users increases, the performance of the model generally deteriorates. This suggests that while adding more fake users tends to reduce the effectiveness of the lists generated, managing a higher number of fake users becomes increasingly difficult.

\subsection{RQ3: Influence of Fake User Interactions on Popular and Unpopular Items}

We investigated whether popular and unpopular items were favoured in recommendation lists by analysing the percentage of the top 20 items recommended to each user. Our results show that unpopular items were consistently underrepresented in these lists. This suggests that more users, a wider range of items, or consideration of a larger number of top positions (e.g. top 100 items) may be necessary to gain a better understanding. On the other hand, in the ML-100k dataset, the percentage of popular items in the recommendation lists without any user-specific adjustments is 5.73\%. The introduction of popular users barely affects this percentage (5.68\%), while the inclusion of non-popular users slightly reduces it to 5.45\%.

These results suggest significant opportunities for future research, such as focusing on specific categories of items to either improve or reduce recommendation performance.

\section{Conclusion}

In this work we investigated the impact of fake users on real users. These fake users can have random interactions, interact with popular or unpopular items, and are only added to the training set at different percentages of the total dataset. The results showed that although the standard metrics were not significantly affected, with random perturbations causing the most significant degradation in performance, the output lists generated under these perturbations were significantly different from the standard lists trained without any perturbations. These differences, measured using ranking list sensitivity metrics, in particular Jaccard and FRBO, showed that in the case of MovieLens about half of the list elements were shared, whereas in the case of Foursquare almost no elements were considered. Furthermore, the proportion of popular and unpopular items in recommendations for real users was not affected by the presence of fake users.

This study opens up future research directions in a number of ways. First, it would be valuable to compare the number of recommended items - categorised as popular, unpopular and genre-specific - using a standard training model with those generated by a model trained on fake users. This comparison could reveal better significant differences in recommendation patterns. Second, the creation of a set of fake users could allow to systematically elevate or downgrade certain categories over time. Third, studying datasets with shorter interaction sequences, such as those from Amazon \cite{hou2024bridging}, could provide new insights into user behaviour and recommendation effectiveness. Finally, research should focus on building resilient models for these types of perturbations: the solution could lie in different training strategies\cite{petrov2022effective}, robust loss functions \cite{bucarelli2023leveraging, wani2024learning}, or different optimisation objectives \cite{bacciu2023integrating}.

\bibliographystyle{ACM-Reference-Format}
\bibliography{citations}

\end{document}